\documentclass{sig-alternate-2013}
\usepackage{amsmath}
\usepackage{amsfonts}
\usepackage{amssymb}
\usepackage{graphicx}
\usepackage{epstopdf}
\usepackage{caption}
\usepackage{listings}
\usepackage{paralist}
\usepackage{tikz}
\usepackage[ruled,commentsnumbered,linesnumbered]{algorithm2e}
\usepackage{lipsum}
\usepackage{hyperref}
\usepackage{url}
\usepackage{multicol}
\usepackage{blindtext}
\usepackage{array}
\usepackage{floatrow}
\usepackage{caption}
\usepackage{lipsum}
\usepackage{wrapfig}
\usepackage{grffile}
\usepackage{xcolor}
\usepackage{color}
\usepackage{algorithmic}
\usepackage{booktabs}
\usepackage{multirow}
\usepackage{alltt}

\usepackage{listings}
\usepackage{color}

\definecolor{dkgreen}{rgb}{0,0.6,0}
\definecolor{gray}{rgb}{0.5,0.5,0.5}
\definecolor{mauve}{rgb}{0.58,0,0.82}

\hypersetup{
    bookmarks=true,         
    unicode=false,          
    pdftoolbar=true,        
    pdfmenubar=true,        
    pdffitwindow=false,     
    pdfstartview={FitH},    
    pdftitle={My title},    
    pdfauthor={Author},     
    pdfsubject={Subject},   
    pdfcreator={Creator},   
    pdfproducer={Producer}, 
    pdfkeywords={keyword1} {key2} {key3}, 
    pdfnewwindow=true,      
    colorlinks=true,       
    linkcolor=blue,          
    citecolor=blue,        
    filecolor=magenta,      
    urlcolor=blue           
}

\input{prelude1}

\begin{document}

\title{StarL: Towards a Unified Framework for Programming, Simulating and Verifying Distributed Robotic Systems}



\maketitle

\begin{abstract}
We developed StarL as a framework for programming, simulating, and verifying distributed systems that interacts with physical processes. 
StarL framework has (a) a collection of distributed primitives for coordination, such as mutual exclusion, registration and geocast that can be used to build sophisticated applications, 
(b) theory libraries for verifying StarL applications in the PVS theorem prover, and 
(c) an execution environment that can be used to deploy the applications on hardware or to execute them in a discrete event simulator.
The primitives have (i) abstract, nondeterministic specifications in terms of invariants, and assume-guarantee style progress properties, (ii) implementations in Java/Android
that always satisfy the invariants and attempt progress using best effort strategies. 
The PVS theories specify the invariant and progress properties of the primitives, and have to be appropriately instantiated and composed with the application's state machine to prove properties about the application.
We have built two execution environments: one for deploying applications on Android/iRobot Create platform and a second one for simulating large instantiations of the applications in a discrete even simulator. 
The capabilities are illustrated with a StarL application for vehicle to vehicle coordination in a automatic intersection
that uses primitives for point-to-point motion, mutual exclusion, and registration. 
\end{abstract}

\setlength{\textfloatsep}{6pt}

\section{Introduction}
\label{sec:intro}

Programs that monitor and control physical processes over a network are becoming common in robotics~\cite{DBLP:journals/arobots/LindseyMK12,kivaFORBES,DBLP:journals/arobots/TurpinMMK14}, smart homes~\cite{export:157701}, and flexible manufacturing~\cite{DBLP:journals/ijmms/LucasT03}. An execution of such a program (consider, for example, a robotic swarm~\cite{kivaFORBES}) is not determined by the underlying computing stack alone, but it also depends on the physical and the network environment. Therefore, to support useful formal reasoning about these types of programs the semantic framework should account for the nondeterminism arising from concurrency, message delays, failures, and the uncertainties in the physical world. 
In swarm robotics, for instance, there is a big gap between the semantics of the models used for proving theorems and the real environment in which the systems run. 
The former is typically a synchronous network without message delays, collision-free physics, etc., while actual implementations use ad hoc strategies for dealing with message losses, noise, and obstacle avoidance, etc. 

We are developing the StarL framework~\cite{newstarl} to bridge this gap by providing a nondeterministic programming abstraction that is sufficiently detailed for proving theorems about reliability and performance of the system, and yet does not overwhelm the application developers. 
%
The core of StarL is a collection of {\em primitives\/}---mutual exclusion, point-to-point motion, leader election, geocast, set-agreement, and many more---that are useful for building distributed robotic applications ({\em StarL applications\/}). 
Each primitive has a hardware-independent, abstract, and nondeterministic specification and an open source implementation in Java. 
The motion-related primitives also have platform specific implementations.
StarL applications running on robots are written in Java and use these primitives to accomplish sophisticated coordination tasks. 
Example StarL applications we have built include a distributed search in which a collection of robots coordinate to find tagets in a building, a light painting application in which a given diagram's outline is painted collaboratively by a collection of robots, and a traffic intersection coordination protocol (see Section~\ref{sec:traffic}). The primitives not only allow us to develop verifiable code for distributed robotics, but offer easy code reuse and maintenance. 

The second component of StarL is the {\em StarL PVS library\/} of theories modeling the abstract specifications of the primitives in the language of the PVS theorem prover~\cite{PVS96:CAV}. These specifications are nondeterministic and have two parts. The first part asserts an invariant property of the primitive and the second part asserts an assume-guarantee style progress property~\cite{Henzinger:2000:DRP:602902.602958}. 
For example, an abstract specification of the mutual exclusion primitive is parameterized by a set of identifiers for participating processes, and the identity of the critical section(s). The specification states (a) no two participating processes occupy the critical section simultaneously (invariant), and (b) that if there exists a time bound within which acquired critical sections are released (assumption) then there exists a time bound within which any requesting process gains access to the critical section. 
In this paper, we show how the PVS theorem prover can be used to develop theories and to verify key invariant properties of StarL applications using the above mentioned primitive theories and their nondeterministic specifications\footnote{Progress properties are verified by composing assumptions and guarantees but this will be the topic of a future paper. 
}.

The third component of StarL is a collection of  {\em StarL execution environments\/} that can be used to deploy the applications on actual hardware or in a simulator. 
We have built two execution environments for StarL: one for deploying Applications on our Android/iRobot platform and a second one for simulating large instances of the applications in a discrete even simulator. 
To our knowledge, StarL is the first framework that enables the creation of verified software for distributed robotic systems.

We provide an overview of StarL in Section~\ref{sec:overview}. Then we illustrate application development in StarL with one detailed example in Section~\ref{sec:traffic}. The PVS translation and verification of this application in Section~\ref{sec:starlpvs}. 
Section~\ref{sec:primitives} describes some of the other StarL primitives.
Section~\ref{sec:sims} describes the execution environments and the simulator. 
Finally, we discuss related work in Section~\ref{sec:related} and conclude in Section~\ref{sec:conc}.

\section{Overview of StarL}
\label{sec:overview}

This work builds up on the work of Zimmerman's master's thesis~\cite{AZimmerman:Masters2013,DuggiralaJZM12}. 
We concretize the concept of primitives, build the connection to PVS, and develop several new applications. 

\paragraph{Primitives}
Deterministic abstractions are easier to program than nondeterministic ones. Programs for a distributed robotic system, however, have to deal with nondeterminism from communication, dynamics, and failures. We make the choice of exposing these nondeterminisms to the programmer through the StarL primitives. 
We ameliorate the loss of determinism by making the primitives {\em uniform\/} in the following way:
The StarL architecture defines a special set of write-one, read-many objects that are stored in a part of the heap called the {\em global variable holder (gvh)\/} for each participating process. A StarL program interacts with a primitive by invoking a set of {\em StarL functions\/} that access these {\em StarL objects\/} in {\em gvh}. 
For example, the StarL $\starl{Mutex}$ primitive implements a distributed mutual exclusion algorithm that allows 
 fixed set of processes $\mathit{PList}$ to access an object in a mutually exclusive fashion. A StarL application uses this primitive as follows:
\begin{lstlisting}[mathescape,xleftmargin=4.0ex]
$\starl{mux = Mutex(\mathit{id,PList})}$; //exclusive with PList			
$\starl{mux.do\_mutex(myreq)}$;   	
while ($\neg \starl{mux.crit}$ && $\neg \starl{mux.failed}$)
	// wait 
if ($\starl{mux.crit}$)
	{ 
		// use
		$\starl{mux.release(myreq)}$;
	}
\end{lstlisting}
The variables $\starl{mux.crit}$ and $\starl{mux.failed}$ in process $i$'s {\em gvh\/} are written by the mutual exclusion algorithm to indicate that $i$ does or does not have access to the requested set $\mathit{myreq}$, or whether the mutex algorithm has failed. 
Similarly, the StarL $\starl{MotionControl}$ primitive implements a path planning and motion control primitive
that interfaces with the low-level motors and actuators and enables the robot running the process to move towards a target $\mathit{t}$ 
while avoiding a region $\mathit{A}$. An application uses this primitive as follows:

\begin{lstlisting}[mathescape,xleftmargin=4.0ex]
$\starl{mc = MotionControl(...)}$; 	
$\starl{mc.do\_move(t,A)}$;   	
while ($\neg \starl{mc.motionflag}$ && $\neg \starl{mc.failed}$)
	// wait 
if ($\starl{mc.motionflag}$)
	{ 
		// reached
	}
\end{lstlisting}
The variables $\starl{mc.motionflag}$ and $\starl{mc.failed}$ in process $i$'s {\em gvh\/} are updated by the motion control algorithm to indicate to the application if $i$ has arrived at the target or whether the motion control has failed. 

%
\paragraph{Verification}
Each primitive not only has a Java implementation but also has formal specifications that state their key invariants and assume-guarantee style progress properties. 
These properties are written in the language of the PVS theorem prover~\cite{PVS:language}. In  Section~\ref{sec:PVSoverview}, we describe how these primitive theories are composed with the specification of the application to create complete PVS theories that can then be verified using the PVS prover. In this paper, we focus on the invariant properties which are proved inductively using the Timed Automaton Library for PVS~\cite{archer:amai2001,ALLMU:memocode06}. The progress properties involve compositional assume-guarantee proofs that are commonly used in the proof of self-stabilizing algorithms~\cite{dolevBook00} and will be the subject of a future paper.

\begin{figure}
\centerline{\includegraphics[scale=0.31]{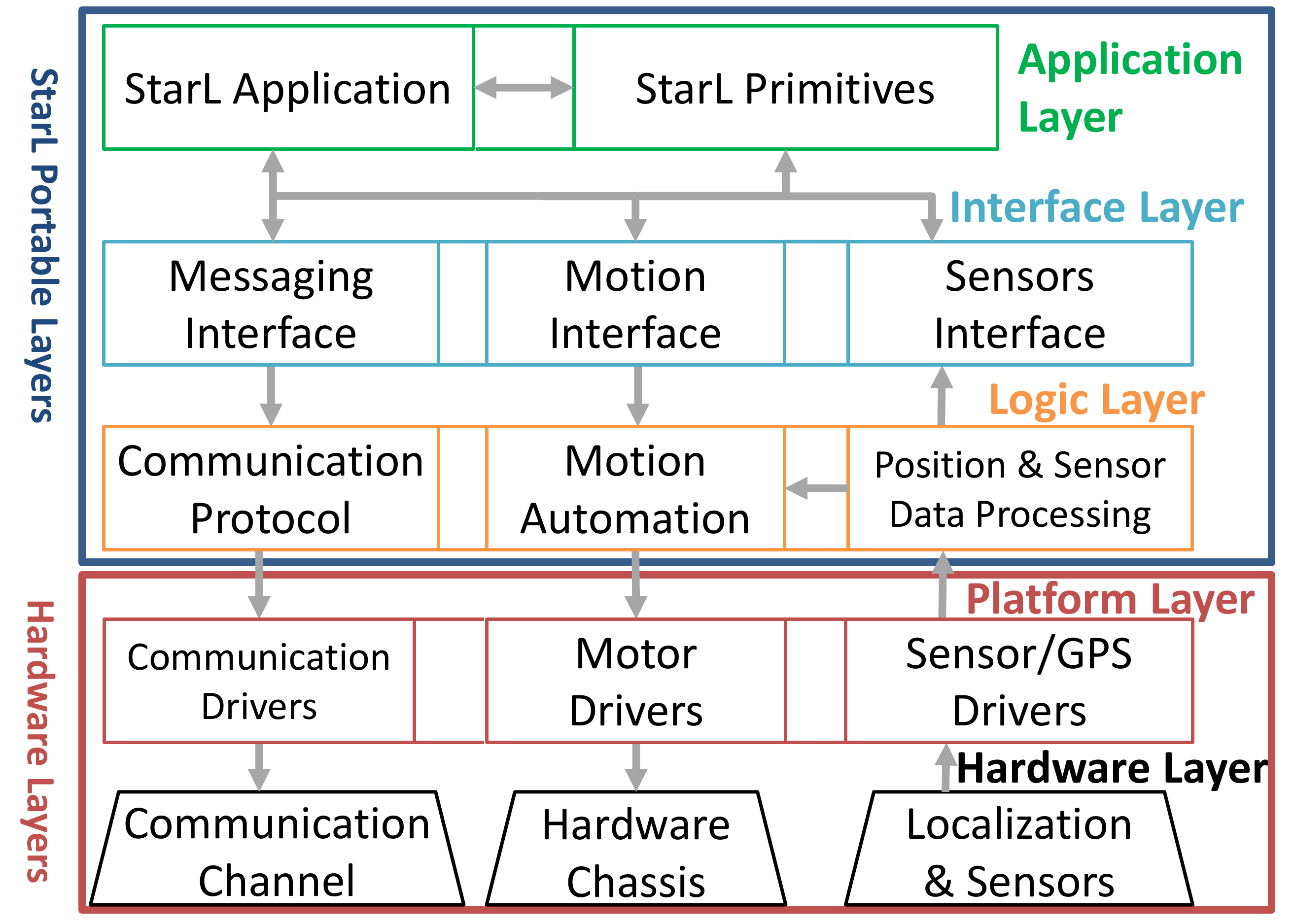}}
\caption{StarL Architecture.}
\label{fig:architecture}
\end{figure}

\paragraph{Architecture}
A robot interacts with the physical environment through sensors and motors. It also interacts with other robots through the communication channels. All of these constitute the {\em execution environment\/} of a StarL application and are organized into five layers as shown in Figure~\ref{fig:architecture}. The bottom two layers are hardware platform specific and the top three layers are portable (see Section~\label{sec:hw} for more details). 
For deploying StarL Applications on our Android/iRobot platform, the platform layer implements the functions for controlling motion of the iRobot Create robots, wireless communication, and for reading data from the OptiTrack indoor positioning system.
For simulating the applications, the platform layer is simulated using models of robot motion and communication channels.
The logic layer wraps the low level methods into high level methods that will be provided to construct the interface layer. 
The interface layer constitutes the global variable holder {\em (gvh)\/} and the various StarL functions to pass data in and out of rest of the stack.
It is an organized collection of all underlying StarL functionality. Through the interface layer, applications may access each part of the framework. 
The top layer is the application layer. This includes StarL primitives (Section~\ref{sec:primitives}) as well as the StarL applications. 
The StarL primitives are constructed using methods from the interface layer. The StarL application uses both interface layer methods and StarL primitives to accomplish more complicated tasks. 
\section{Automatic Intersection}
\label{sec:traffic}

We discuss the key facets of StarL with an automatic intersection application. Automatic intersection protocols that exploit vehicle to vehicle (V2V) communication have been proposed at various levels of detail in the context of smart cities and autonomous cars~\cite{DBLP:journals/tits/HafnerCCV13,JMS:ICDCS2010}. We use a toy version of this application to illustrate improvements in programmability and verifiability with StarL. 

\begin{figure}
\centerline{\includegraphics[scale=0.33]{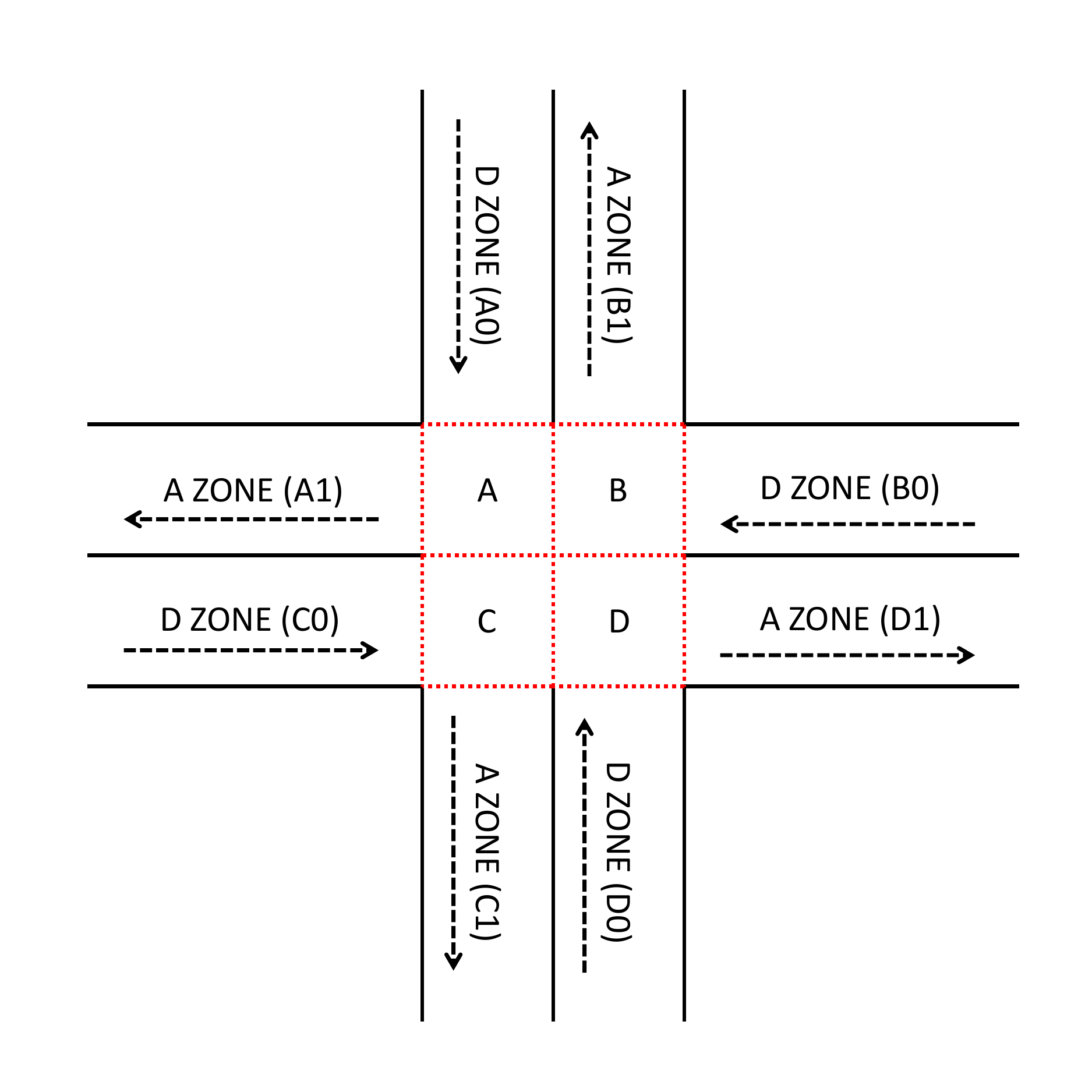}}
\caption{The four-way automatic intersection.}
\label{fig:fourway}
\end{figure}

\paragraph{Automatic intersection layout}
Consider a four-way, double-lane, intersection that will be navigated by autonomous robotic vehicles through communication (see Figure~\ref{fig:fourway}). 
Each vehicle arrives at one of the {\em arrival zones\/} $\mathit{A0, B0, C0, D0}$ with a designated {\em departure zone\/} $\mathit{A1, B1, C1, D1}$. It coordinates with the other vehicle according to a {\em intersection coordination protocol (ICP)\/} and proceeds to move through a sequence of {\em critical zones\/} $\mathit{A, B, C, D}$ following certain right-hand traffic rules (e.g., no backing or U-turns). 
For example, a vehicle with source destination pair $\mathit{(A0, D1)}$ will have the path $\mathit{A0, A, C, D, D1}$.
The requirements from the system are:
\begin{enumerate}[(a)]
\item {\sf (traffic\_safety)\/} No two vehicles occupy the same critical zone at the same time. 
\item {\sf (traffic\_progress)\/} There exists a time-bound within which every approaching vehicle departs. 
\end{enumerate}
We would also like the protocol to permit concurrent safe traversals. For examples, vehicles with paths $\mathit{A0, A, A1}$ and $\mathit{D0, D, B, B1}$ should not block each other. 

\subsection{Intersection Coordination using StarL}
\label{sec:ICP_implement}
A protocol for intersection works as follows: the participating vehicles agree on the set of participants, 
then they request access to the sequence of zones needed for traversal in the intersection from the set of agreed-upon participants; 
once they have access to the entire sequence, they start traversing; when a zone is crossed it is released. 
For the sake of simplicity, in this presentation we assume that processes do not fail and robots do not get stuck. 

Figure~\ref{fig:gen_alg} shows the code for implementing this ICP using StarL primitives and Figure~\ref{fig:java} shows the actual Java implementation.
Each vehicle participating in the coordination runs an instance of this protocol with the same identifier $\mathit{xid}$ that 
uniquely identifies the intersection. 
For the process at vehicle $i$, the local variable $\mathit{plist}_i$ is a list of identifiers of participating process initialized to the emptylist. 
The local variable $\mathit{myseq}_i$ is a list of zones;
it is initialized to the sequence of zones that $i$ must traverse to go from its current position ($\starl{mypos}$) to its destination.
Here, $\starl{mypos}$ is a StarL variable in the {\em gvh\/} storing the position of the vehicle and is updated by the location sensors. 
This protocol uses two StarL primitives called $\starl{Registration}$ and $\starl{Mutex}$.
More details about these primitives, their interfaces, and the conditional guarantees they provide are described in Section~\ref{sec:primitives}.
In brief, $\starl{Registration}$ allows a set of processes in a neighborhood to agree on a subset that contains participating processes; 
$\starl{Mutex}$ allows mutually exclusive access to one or a set of shared resources.
$\starl{reg}$ and $\starl{mux}$ are instances of these primitives with the identifier $\mathit{xid}$.
Finally, the $\mathit{loc}$ variable, of the enumerated type, is initialized to the value $\texttt{S0}$.

The protocol waits in a loop until $\mathit{loc}$ becomes $\texttt{S0}$.
If $\mathit{loc}$ is $\texttt{S0}$ then the $\starl{do\_register()}$ function is invoked to start the registration process
and $\mathit{loc}$ is set to $\texttt{reg\_wait}$. 
If and when the registration process returns successfully, the StarL variable $\starl{reg.rList}$ is set to a non-null value. 
From \texttt{reg\_wait}, the process moves to \texttt{mutex\_wait} only if registration completes ($\starl{reg.rList}$ nonempty) and in that case 
the list is copied to local variable $\texttt{plist}$ and $\starl{do\_mutex}$ is invoked to obtain exclusive access to the sequence of zones $\mathit{mid(myseq_i)}$ (except the first and the last) from the processes in $\mathit{plist}$.
If and when the mutex process returns successfully, the StarL variable $\starl{mux.crit}$ is set to $\texttt{true}$. 
From \texttt{mutex\_wait}, the process moves to \texttt{move\_wait} only if mutex returns successfully 
($\starl{mux.crit}$ true) and in that case $\starl{do\_move}$ is invoked which sends from $\mathit{plist}$
a sequence of points to $\starl{Motion control}$. 
In \texttt{move\_wait}, when the vehicle traverses the zone $\mathit{plist[1]}$ and reaches $\mathit{plist[2]}$, the zone $\mathit{plist[1]}$ is removed from the list and $\starl{release}(\texttt{plist[1]})$ is called to release that zone to the mutual exclusion. 
When the vehicle $i$ reaches its destination zone, the $\mathit{loc}$ is changed to \texttt{S1}.

\begin{figure}
\lstset{escapeinside={(*@}{@*)}}
\begin{lstlisting}[mathescape,xleftmargin=4.0ex]
plist: List[PIDS] := {};       (*@\label{ln:createvC}@*) 
myseq: List[Zones] := path($\starl{mypos}$,dest)         (*@\label{ln:getwantedC}@*) 
reg = $\starl{Registration(\mathit{xid})}$;  		(*@\label{ln:getservC}@*) 
mux = $\starl{Mutex(\mathit{xid})}$;			
loc enum {S0,reg_wait,mutex_wait,move_wait,S1} := S0		(*@\label{ln:enumC}@*) 
while (state != done)							(*@\label{ln:startC}@*) 
	switch case state
		S0: loc = reg_wait; $\starl{reg.do\_register()}$;  
		reg_wait: if $\starl{reg.rlist}$ != null then 
			state = mutex_wait;
			plist = $\starl{reg.rList}$;			(*@\label{ln:plistC}@*) 
			$\starl{mux.do\_mutex(mid(myseq),plist)}$;    (*@\label{ln:S0C}@*) 
		mutex_wait: if $\starl{mux.crit}$ = myseq then	 (*@\label{ln:mutexwaitC}@*) 
			state = move_wait;
			$\starl{do\_move(plist)}$
		move_wait: if $\starl{pos} \in$ seq[2] then      (*@\label{ln:reachedC}@*) 
			$\starl{mux.release(seq[1])}$; myseq = tail(myseq);		(*@\label{ln:releaseC}@*) 
			if myseq = [dest] then state := S1		 (*@\label{ln:moveoutC}@*) 
		S1:	//done
\end{lstlisting}
\caption{General Algorithm.}
\label{fig:gen_alg}
\end{figure}



\subsection{Java Implementation of ICP}
\label{sec:javaimp}

Figure~\ref{fig:java} shows a fragments of Java implementation of ICP that highlights the usage of the $\starl{Mutex}$ and $\starl{Motion control}$ primitives. 
Line~\ref{ln:enum} in Figure~\ref{fig:java} (corresponds to line~\ref{ln:enumC} in Figure~\ref{fig:gen_alg}) enumerates the program locations. 
Line~\ref{ln:createv} (corresponds to Line~\ref{ln:createvC} in Figure~\ref{fig:gen_alg}) creates the variable $\mathit{plist}$ to store the list of vehicles that will be returned from in the $\starl{Registration}$ primitive. Then it creates a variable ($\mathit{myseq}$) to hold the list of wanted zones. It is initialized by computing the sequence of critical zones plus the departure zone the vehicle needs to go through this intersection. 
Line~\ref{ln:getserv} (line~\ref{ln:getservC}) creates a instance of the registration primitive in the {\em gvh} using the intersection ID. 
Similarly, a instance of the mutual exclusion primitive is created in the {\em gvh} in the next line.

When access is granted by \starl{Mutex} (line~\ref{ln:motion}), it sends the motion command $\starl{do\_move}$ and changes $\mathit{loc}$ to $\mathit{move\_ wait}$. The the Java code, $\texttt{gvh.plat.moat}$ refers to the motion automation that controls the movements of the robot. 
The vehicle stops when it has either reached the neighborhood of the destination or failed. In this application, since we want to ensure safety, the program is interrupted if collision is detected. Therefore, the vehicle stops if and only if it has reached the neighborhood of the destination (line~\ref{ln:reached}, (line~\ref{ln:reachedC})). Then, line~\ref{ln:release}, (line~\ref{ln:releaseC}) releases($\starl{release}$) the previous critical zone. Line~\ref{ln:moveout},(line~\ref{ln:moveoutC}) moves ($\starl{do\_move}$) to the next critical zone in $\mathit{myseq}$. Additionally, location is changed to to S\_1 if there are no more zone in $\mathit{myseq}$. 

Line~\ref{ln:s1} waits until the vehicle has reached the neighborhood of the departure zone, then the last critical zone is released and the $\starl{unRegister}$ method is called. Line~\ref{ln:shutdown} freezes the robot, preventing any more motion command to be executed. 
Line~\ref{ln:sleep} makes the execution of the while loop wait so that the states are updated once roughly every 100 milliseconds.
\begin{figure}[ht!]
\lstset{escapeinside={(*@}{@*)}}
\begin{lstlisting}[firstnumber=1,xleftmargin=4.0ex]
	private enum Location { (*@\label{ln:enum}@*) 
		S_0, REGISTER_WAIT, MUTEX_S, MUTEX_WAIT, MOVE_WAIT, S_1, DONE, ...
	};
	LinkedList<ItemPosition> plist;	(*@\label{ln:createv}@*) 
	LinkedList<ItemPosition> myseq = getMyseq();			
	RegPrim Reg = new Registration(gvh,I_ID); (*@\label{ln:getserv}@*) 
	MutexPrim Mutex = new M_Mutex(gvh, I_ID); 

@Override
public List<Object> callStarL() {(*@\label{ln:start}@*) 
	while(location != DONE) {	
		switch(location) {
// implementation of some locations not shown
		case REGISTER_WAIT:
			if(Reg.getList() != null){
				plist=Reg.getList();	(*@\label{ln:plist}@*) 
				Mutex.do_mutex(myseq,plist);   (*@\label{ln:S0}@*) 
				location = Location.MUTEX_WAIT;
			}
		break;			
		case MUTEX_WAIT: (*@\label{ln:mutexwait}@*) 
			if(Mutex.od_mutex()){
				gvh.plat.moat.doMove(currentDestination); (*@\label{ln:motion}@*) 
				location = Location.MOVE_WAIT;
			}
		break;		
		case MOVE_WAIT:{ (*@\label{ln:movewait}@*) 
			if(!gvh.plat.moat.inMotion) 	 (*@\label{ln:reached}@*) 
				Mutex.release(CSname(preDestination));  (*@\label{ln:release}@*) 	
				preDestination = new ItemPosition(currentDestination); (*@\label{ln:predest}@*) 
				myseq.remove()
				if(!myseq.isEmpty()){ 
					currentDestination = (ItemPosition)myseq.peek();
					gvh.plat.moat.doMove(currentDestination);		(*@\label{ln:moveout}@*) 
				}; 
				else{
					location = Location.S_1;
				}
			}
		break;	
		case S_1:
			if(!gvh.plat.moat.inMotion) { (*@\label{ln:s1}@*) 
				Mutex.release(CSname(preDestination));
				Reg.unRegister();
				preDestination = null;
				gvh.plat.moat.motion_stop();	(*@\label{ln:shutdown}@*) 
				location = Location.DONE;
			}
		break;
		case DONE: (*@\label{ln:donet}@*) 
			break;
		}
	sleep(100);  (*@\label{ln:sleep}@*) 
	}
}
\end{lstlisting}
\caption{ICP Java Implementation.}
\label{fig:java}
\end{figure}
\subsection{StarL PVS Library}
\label{sec:starlpvs}

The StarL application code and the primitives can be translated to the PVS theorem prover's language of high order logic, for rigorously proving safety and progress properties with appropriate environmental assumptions. 
Figure~\ref{fig:pvs} shows the key part of the PVS theory specifying a system running the ICP application. It defines the semantics of the system in terms of a timed automaton~\cite{TIOAmon}. Although the pseudo code of Figure~\ref{fig:gen_alg} and its Java implementation Figure~\ref{fig:java} are for an individual processes, this PVS theory (together with its supporting and importing theories) specify the behavior of the entire system with arbitrarily number of asynchronously evolving processes.

\subsection{Overview of the PVS Theories}
\label{sec:PVSoverview}
The theory uses the TAME library~\cite{archer98tame,DAEM08} for modeling timed automata in PVS. The body of the theory defines the states, the start states, the actions, and the transitions of the automaton---a special action $\mathit{dt}$ models the passage of time. By importing the $\mathit{time\_machine}$ theory with these parameters (Line~\ref{pvsln:importmutex}), the generic timed automaton theory is instantiated and that gives the instances of the relevant definitions and theorems (e.g., the notion of reachable states, invariants, and inductive proof rules) for this model. 

The interface and the implementation of each StarL primitive is defined in separate parameterized PVS theories such as $\pvs{Mutex_decls}$ and $\pvs{Registration_decls}$. 
The $\pvs{Traffic_decls}$ theory imports appropriate instances of these theories. In order to define a timed automaton that is the composition of several primitives (\starl{Mutex} and \starl{Registration}), in this paper we develop an approach for compositional modeling of timed automata in PVS. For the sake of brevity, the theory presented here excludes the registration process and we drop the parts related to timing behavior. 

Line~\ref{pvsln:states} defines the state components of this automaton.
time $\mathit{loc}$ and $\mathit{myseq}$ variables correspond to the variables with the same name in Figure~\ref{fig:gen_alg}. However, notice that here they are arrays indexed by the process (PID), i.e., they model the location and the sequence of zones for all the processes in the system. 
The variable $\mathit{timer}$ is a global clock used to prove time-bound properties\footnote{The details of timing analysis will be presented in a future paper.}
The $\mathit{timer\_move}$ variable is a stopwatch that tracks, for each vehicle, the duration of physically traversing zones. 
The state component $\mathit{mux}$ is the state of the imported Mutual exclusion primitive.

The $\mathit{action}$ datatype defines the names and types of all the state transition. 
The $\mathit{enabled}(a,s)$ predicate defines whether the action $a$ {\em can\/} occur in state $s$ and the $\mathit{trans}$ function function defines the post-state of $a$ occurring at $s$. 
$\mathit{dt}$ models the progress of real time; 
$\mathit{do\_mutex}(i,Z)$ models the $i^{th}$ process requesting the set of zones $Z$;
$\mathit{od\_mutex}(i,Z)$ models successful completion, i.e., the mutual exclusion primitive granting $i$ access to $Z$; and 
$\mathit{release}(i,Z)$ models process $i$ crossing a zone and releasing it to the mutual exclusion primitive. 
Note the enabling condition for completing Mutex ($\mathit{od\_mutex}$, Line~\ref{pvsln:od_mutex_enabled}): it is a conjunction of a condition from the ICP and a condition from Mutex; this captures composition of the two automata. 
Similarly in Lines~\ref{pvsln:do_mutex_trans}-\ref{pvsln:od_mutex_trans}, the transition function for the two actions are combining the transitions of ICP and Mutex.

\begin{figure}[h]
\centering
  {
\begin{lstlisting}[language=pvsNums,xleftmargin=4.0ex]
Traffic_decls : THEORY BEGIN
	IMPORTING Mutex_decls[PIDS,Zones]	$\label{pvsln:importmutex}$

	sd: array[PIDS $\rightarrow$ (Valid_sd?)] 	

	states: TYPE = [#	loc: array[PIDS $\rightarrow$ Locations], $\label{pvsln:states}$	
		myseq: array[PIDS $\rightarrow$ ZoneList],
		timer_move: array[PIDS $\rightarrow$ nonnegreal],
		timer: nonnegreal,
		mux: Mutex_decls.states #]  $\label{pvsln:muxs}$	

	actions: DATATYPE BEGIN	$\label{pvsln:actions}$
		dt(delta_t:nonnegreal): dt?
		do_mutex(i:PIDS,RS:Zoneset): do_mutex?
		od_mutex(i:PIDS,RS:Zoneset): od_mutex?
		release(i:PIDS,RS:Zoneset):release?
	END actions

	enabled(a:actions, s:states):bool = CASES a OF	$\label{pvsln:enabled}$
		dt(delta_t):  ...
		do_mutex(i,RS): loc(i,s) = S0 AND RS = list2set(Path(sd(i))),
		od_mutex(i,RS): loc(i,s) = mutex_wait AND $\label{pvsln:od_mutex_enabled}$
			Mutex_decls.enabled(od_mutex(i,RS),mux(s)),
		release(i,RS): loc(i,s) = move_wait AND RS = car(myseq(i,s)) AND (NOT myseq(i,s) = null)	
		ENDCASES

	trans(a:actions, s:states):states = CASES a OF
		dt(delta_t): s WITH ...
		do_mutex(i,RL): s WITH 
			[loc:= loc(s) WITH [(i):= mutex_wait],
				mux:= mutex_decls.trans(do_mutex(i,RL),mux(s))], $\label{pvsln:do_mutex_trans}$
		od_mutex(i,RL): s WITH 								$\label{pvsln:od_mutex_trans}$
			[loc:= loc(s) WITH [(i):= move_wait],
				mux:= mutex_decls.trans(od_mutex(i,RL),mux(s))],
		release(i,RL): IF myseq(i,s) = null THEN s WITH 
			[loc:= loc(s) WITH [(i):= S1] ] ELSE s WITH 
			[myseq:= myseq(s) WITH [(i):= cdr(myseq(i,s))],
				timer_move := timer_move(s) WITH [(i):= 0],
				mux:= Mutex_decls.trans(release(i,RL),mux(s))]
			ENDIF
		ENDCASES

	IMPORTING time_machine[states,actions,enabled,trans,start] 
END Traffic_decls
\end{lstlisting}
	}
  \caption{PVS theory for (part of) ICP.}
  \label{fig:pvs}
\end{figure}

\subsection{Proving Theorems about ICP}
\label{sec:PVSoverview}

The above PVS theory defines a timed automaton model and its semantics for a system with an arbitrary number of processes executing the ICP which in turn involves those processes participating in the $\starl{Mutex}$ primitive. We give a sketch of the key invariants that are used to prove safety of ICP using the PVS theorem prover.
The \starl{Mutex} primitive is not presented in detail in this paper. It involves a set of processes and has a key component $\mathit{critset:[PIDS \rightarrow Zoneset]}$ that records the (possibly empty) set of zones that each process has exclusive access to.
Its key invariant property is stated in Line~\ref{pvsln:mutexinvs} of Figure~\ref{fig:pvsinvs}.
It asserts that at any reachable state $s$ of the \starl{Mutex} primitive, for any pair of processes $i$ and $j$, $\mathit{crit\_set(i,s)} \cap \mathit{crit\_set(j,s)} = \emptyset$. 

The next inductive invariant (Line~\ref{pvsln:invlistcrit}) states what we found to be the key property needed for proving safety of ICP:
it asserts that for any process $j$, 
(a) if $j$ is in $S0$, then $\mathit{myseq}(j,s)$ is the list of zones from its current position to the destination, 
(b) if $j$ is in $\mathit{mutex\_wait}$, then $\mathit{myseq}(j,s)$ is same as the list in (a) and it has requested to $\starl{Mutex}$ exclusive access to all the elements in this list, and
(c) if $j$ is in $\mathit{move\_wait}$, (i.e., $\starl{Mutex}$ has completed and $j$ is moving), 
then $\mathit{myseq}(j,s)$ is a subset of $\mathit{crit\_set}(j,s)$.
Using this invariant, the main safety invariant (Line~\ref{pvsln:safety}) is proved: it states that for any two processes that are moving, the occupy different zones. 

\begin{figure}[h]
\centering
  {
\begin{lstlisting}[language=pvsNums,,xleftmargin=4.0ex]
Inv_mux_safety(s):bool = FORALL (i,j): NOT (i = j) IMPLIES disjoint?(crit_set(i,s),crit_set(j,s))
lemma_mux_safety: LEMMA FORALL (s): reachable(s) => Inv_mux_safety(s); $\label{pvsln:mutexinvs}$

Inv_list_crit(s):bool = FORALL (j): 
   (loc(j,s) = move_wait IMPLIES subset?(list2set(myseq(j,s)),crit_set(j,s))) AND
   (loc(j,s) = mutex_wait IMPLIES list2set(myseq(j,s))= req_set(j,s) AND myseq(j,s) = Path(sd(j))) AND
   (loc(j,s) = S0 IMPLIES myseq(j,s) = Path(sd(j)))
lemma_list_crit: LEMMA FORALL (s): reachable(s) => Inv_list_crit(s); $\label{pvsln:invlistcrit}$

Inv_safety(s):bool =  FORALL (i,j): NOT (i = j) AND loc(i,s) = loc_move_wait AND loc(j,s) = move_wait IMPLIES (NOT (car(myseq(j,s)) = car(myseq(i,s))) OR myseq(j,s) = null OR myseq(i,s) = null) $\label{pvsln:safety}$
lemma_safety: LEMMA FORALL (s): reachable(s) => Inv_safety(s);

% Progress lemmas
lemma_move_entry: LEMMA FORALL (s): FORALL (i): EXISTS (d_1:nonnegreal): reachable(s) AND timer(s) >= d_1 IMPLIES loc(i,s) = loc_move_wait; $\label{pvsln:progmutex}$

lemma_progress: LEMMA FORALL (s): FORALL (i): EXISTS (d_2:nonnegreal): reachable(s) AND timer(s) >= d_2 IMPLIES loc(i,s) = loc_s1; $\label{pvsln:progress}$
\end{lstlisting}
	}
  \caption{PVS theory with key ICP invariants.}
  \label{fig:pvsinvs}
\end{figure}
\section{StarL Primitives}
\label{sec:primitives}

StarL currently includes the implementation of the following primitives: path planning, distributed path planning, geocast, leader election, registration, mutual exclusion, barrier synchronization. In this section, we enumerate the interfaces and specifications for some of them. 

Some of the specifications involve timing properties that are stated with respect to certain intervals defined over real-time. These intervals are defined using constants $d, d_1, d_2$, etc. The role of these constants play in verification are different from the role in implementation. For verifying StarL applications, these constants appear as existentially quantified parameters in the lemma statements (see Line~\ref{pvsln:progmutex}). We assume that certain progress making events happen within some time bound, e.g., delivery of messages, to prove existence of time bounds of other events such as traversal through intersection.
In the actual implementation of the primitives, the progress time bounds may be violated, but they still provide a guideline tuning best-effort strategies.

\subsection{Motion Control}
\label{sec:motioncontrol}
The $\starl{Motion control}$ primitive allows the application to direct the robot towards a specific target point while avoiding a bad region. Both the target and the region are specified in the current coordinate system. The motion completes successfully if the robot reaches a neighborhood of the target while avoiding the bad region and this is indicated to the program.
The interface includes: 
\begin{enumerate}[(a)]
\item $\starl{\langle target, avoid\rangle}$ variable pair in the {\em gvh\/} that stores the target and the bad region, 
\item $\starl{gotopoint}$ function is invoked to set $\starl{target}$ and $\starl{avoid}$,
\item $\starl{motionflag}$ is a boolean variable that is set to {\em done\/} when the motion completes successfully, and it is set to {\em fail\/} to indicate that the lower-level motion controller cannot move the robot to the target.
\end{enumerate}
The following properties summarize the specification of Motion control.

\begin{enumerate}[(a)]
\item {\sf (safety)\/} 
The position of the robot is always outside the region in $\starl{avoid}$.
\item {\sf (progress)\/} If $\starl{motionflag}$ is set to true then the position of the robot is locates near $\starl{target}$. 
\end{enumerate}
%
%
%
The motion control primitive is implemented using lower-level control and path planning algorithms. More details about implementations are provided in Section~\ref{sec:hw}.

\subsection{Geocast and Broadcast}
\label{sec:geocast}
The $\starl{Geocast}$ primitive allows a process to send a message $m$ to all other processes/robots in its' neighborhood $A$; here $A$ is defined by distance, and $d$ is a timing parameter. The following properties specify the behavior of the geocast primitive. The interface includes
\begin{enumerate}[(a)]
\item $\starl{do\_geocast(m,A,d)}$ function to start geocast of message $m$ over area $A$ with timing parameter $d$ (explained below),
\item $\starl{\langle Gcastflag\rangle}$ is a variable in the {\em gvh\/} that indicates that the geocast has completed. 
\end{enumerate}
The following properties summarize the properties of the primitive. If a message $m$ is send through geocast at time $t_0$ then the following hold:
\begin{enumerate}[(a)]
\item {\sf (exclusion)} Any process continuously located outside $A$ during the time interval $[t_0, t_0 + d]$ will not deliver $m$.
\item {\sf (inclusion)} Any process located within $A$ during the time $[t_0, t_0 + d]$ will receive $m$ within $d$ time of the geocast. 
\end{enumerate}

For a robot moving in or out of $A$ during the geocast period, the message may or may not be delivered; but a robot outside $A$ is guaranteed not to receive the message. The implementation of geocast over a wireless network involves details like tagging the message with the location of the originating process before sending, resending messages in the absence of acknowledgments, and dropping the messages based on the receiver's location. Of course, (b) can only be guaranteed under additional assumptions about messages being delivered in a timely fashion.

$\starl{BCast(m,d)}$ or broadcast is a special geocast in which the $A$ defines the entire network. The second condition then requires that all process that are non-faulty over the interval $[t_0, t_0 + d]$ receive the message. 

\subsection{Registration}
\label{sec:reg}
The $\starl{Register}$ primitive solves a set-valued distributed consensus problem for a set of processes to agree on the identity of the participants. If registration completes successfully, then the agreed upon set contains a process's identifier if and only if it is a participating process. 
The interface includes: 
\begin{enumerate}[(a)]
\item $\starl{Register}$ function for creating a register object,
\item $\starl{do\_register}$ function for starting registration,
\item $\starl{\langle rList, ts\rangle}$ pair stores in the {\em gvh\/}; $\starl{rList}$ is the agreed set and $\starl{ts}$ is the time-stamp for when the computation finishes; otherwise $\starl{rList}$ stores a $\mathit{null}$ value. 
\end{enumerate}
The following properties summarize the nondeterministic specification of the Register primitive.

\begin{enumerate}[(a)]
\item {\sf (agreement)\/} 
For any two processes $i$ and $j$ with agreement timestamps ($\starl{ts}$) within $d$ of each other, the corresponding $\starl{rList}$'s are identical.
\item {\sf (soundness)\/} 
For any process $i$, $i$ is contained in $\starl{rList}$ with time stamp $\mathit{ts}$ only if $i$ invoked $\starl{do\_register}$ at most $d_1$ time before $t$. 
\item {\sf (progress)\/} For any process $i$, if $i$ invokes $\starl{do\_register}$ then within at most $d_2$ time registration completes with $i$, that is, $\starl{rList}$ contains $i$. 
\end{enumerate}
%
%
%
The $\starl{Register}$ is implemented using the $\starl{Geocast}$ primitive.
To support multiple registered lists inside an application, each registration object is invoked with an identifier. A registered process may unregister from the list and this essentially restarts a registration process among the remaining processes. The $\starl{rList}$ value can be updated with a new time stamp and in the interim it may be $\mathit{null}$. 

\subsection{Leader Election}
\label{sec:leaderelection}

The $\starl{Election}$ primitive elects a leader and conveys the leader's identity to set of participating processes. If the election fails then the participating processes learn about this as well. The interface includes: 
\begin{enumerate}[(a)]
\item $\starl{Election}$ function for creating an election object; it takes the list of participants as a parameter,
\item $\starl{do\_election}$ function starts the election,
\item $\starl{\langle Leader\rangle}$ stores the identity of the leader in the {\em gvh\/},$\mathit{null}$ if the election is in progress, and $\mathit{fail}$ if the election fails. 
\end{enumerate}
The following properties summarize the nondeterministic specification of the \starl{Election} primitive.
\begin{enumerate}[(a)]
\item {\sf (agreement)\/} 
For any two processes $i$ and $j$ that start election within $d$ time of each other, if $\mathit{Leader}$ is not $\mathit{null}$ or $\mathit{fail}$ for either of the two processes, then 
$\mathit{Leader}$ has identical value for both. 
\item {\sf (soundness)\/} 
For any process $i$, $Leader = i$ only if $i$ invoked $\starl{do\_election}$ at most $d_1$ time before $t$. 
\item {\sf (progress)\/} For any process $i$, if $i$ invokes $\starl{do\_election}$ then within at most $d_2$ time election completes successfully, that is, $\starl{Leader}$ equals a valid identifier. 
\end{enumerate}
%

%
%
Currently, one of the implementations of leader election is based on randomized ballot creation and a second implementation is based on a version of the Bully algorithm~\cite{Coulouris:2011:DSC:2029110}. 

\subsection{Mutual Exclusion}
\label{sec:mutex}
The $\starl{Mutex}$ primitive allows a fixed set of processes to access an object (or a set of objects)in a mutually exclusive fashion. If a process requests multiple objects, then it gains access to all of them at the same time, but it may release them one at a time. The interface includes:
\begin{enumerate}[(a)]
\item $\starl{Mutex}$ function for creating an mutual exclusion object for a list of participating processes and a list of critical sections. 
\item $\starl{do\_election}$ is invoked to request a set of critical sections,
\item $\starl{crit}$ stores in {\em gvh\/} a boolean value indicating whether access to all the requested critical sections have been granted to this process. 
\end{enumerate}
The following properties summarize the specification of the \starl{Mutex} primitive.

\begin{enumerate}[(a)]
\item {\sf (safety)}: For any two processes, the set of critical sections they have access to are disjoint. 
\item {\sf (progress)}: if there exists a time bound $d_1$ within which critical sections are released then there exists a time bound $d_2$ within which any requesting process gains access to its critical section(s). 
\item {\sf (non-interference)}: If no process holds the critical sections being requested by $i$, then $i$ gains access with time $d_3$ ($d_3 \ll d_2$). 
\end{enumerate}
The intersection coordination protocol described in Section~\ref{sec:traffic} uses the \starl{Mutex} primitive. Currently, mutual exclusion is implemented using a modification of Ricart \& Agrawala's algorithm~\cite{Coulouris:2011:DSC:2029110}.

In summary, all the primitives provide a same type of abstraction to the programmer: an set of invariant properties that restrict what the nondeterministic environment can do, and a set of assume-guarantee style progress property. 
The primitives are invoked by calling the interface functions, and 
progress can be detected by reading the appropriate variables in the {\em gvh\/}.

\section{Execution Environments}
\label{sec:sims}

In order to run a StarL application on a hardware platform or inside a simulation environment, it has to be connected with an {\em execution environment\/}. We have developed two execution environments: (1) for running applications on a collection of Android smart phones that control iRobot Create robots and (2) for simulating the applications in a discrete event simulation environment. 
Recall, the execution environment define the lowest two layers of Figure~\ref{fig:architecture} (platform and physical layers), and the rest of the software stack is portable.

\subsection{Deploying Applications on HW Platforms}
\label{sec:hw}
For deploying StarL Applications on our Android/iRobot platform, the platform layer implements the functions for controlling motion of the iRobot Create robots, wireless communication, and for reading data from the OptiTrack indoor positioning system.

The sensor data and the location data from the positioning system are processed, filtered, through the different layers and are recorded in $\starl{currentlocation}$ variables in the {\em gvh\/}.
When the application calls the $\starl{do\_move()}$ in the $\starl{Motion control}$, a motion controller is started that decides when and what command to send to robot while making use of positioning data and sensor data and updating $\starl{motionflag}$ in the {\em gvh\/}. When the controller decides that the robot has to go straight or arc, the platform layer issues the appropriate wheel speed command to the iRobot Create chassis. 
The motion interface also provides underlying motion automation. For example, one can specify a robot's type so that the robot can behave differently when it collides with an object. There are implementations of stop on collision, back away from collision point, discover objects around the initial collision point. 

The message interface provides basic send and receive functions over a Wi-Fi network using our built in protocols. These low-level functions are used to build the $\starl{Geocast}$ primitive.

\subsection{Simulating Applications}
\label{sec:hw}
The same StarL code can also be simulated in a discrete event simulator that we have built. This is useful for testing applications on many virtual robots and without a hardware platform. 
\mitras{resume here.}
The simulator features a custom implementation of the platform layer which directs motion, message, and trace commands into a coordinating thread referred to as the simulation engine. The simulator can execute an arbitrary number of copies of a StarL application code to run and interact simultaneously through simulated messages and robotic platforms.

The StarL simulator allows a developer to run an application under a broad range of conditions and with any number of participating robots. A visualizer displays the current position of each agent and can be extended to display additional application specific information. Even we could now produce a simulation environment same to the real robotic platform, the challenges robots face are realistic. A large set of simulating parameters can be tuned. Message delays, message loss rate, obstacles in the physical environment, robots crash failures and even adversary robots are among the tunable simulation parameters.

Creating the simulation in StarL is simple. Using our simulation template, one need to specify the application (figure~\ref{fig:java}) to simulate along with some simulation parameters. For example, one can simulate ICP with 4 robots, 100 milliseconds average message delay, the obstacles in the physical environment, shown in figure~ref{fig:sim1}. One can also customize the visualizer to display some extra application specific information, such as the state of the robots.

\subsection{The ICP Application}
\label{sec:ICPapp} 

\begin{figure}
\centerline{\includegraphics[scale=0.33]{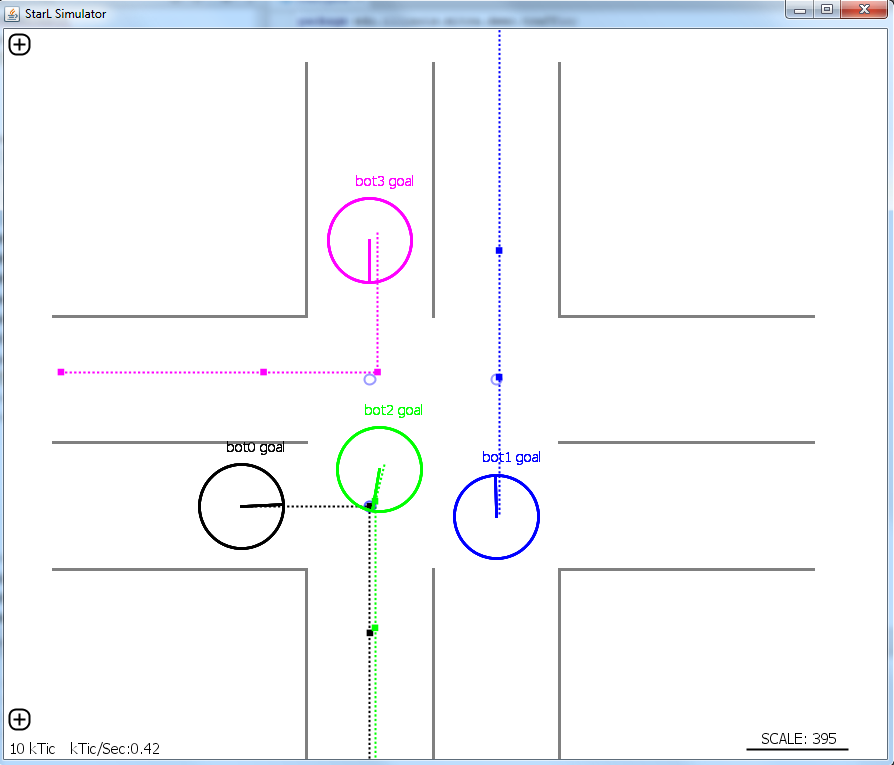}}
\caption{A snapshot of the ICP simulation.}
\label{fig:sim1}
\end{figure}

Screen shot for simulating ICP general solution using four robots is shown in figure~\ref{fig:sim1}. Robot 2 starts in B0, and intends to turn left. Robot 1 starts in D0 and intends to go straight. Robot 0, starting at C0, and robot 3, starting in A0, both intend to turn right. The dotted lines are intended zone sequence.
%
Robot 1 and robot 2 are in the intersection concurrently since their set of critical zones are disjoint. 

\begin{figure}
\centerline{\includegraphics[scale=0.25]{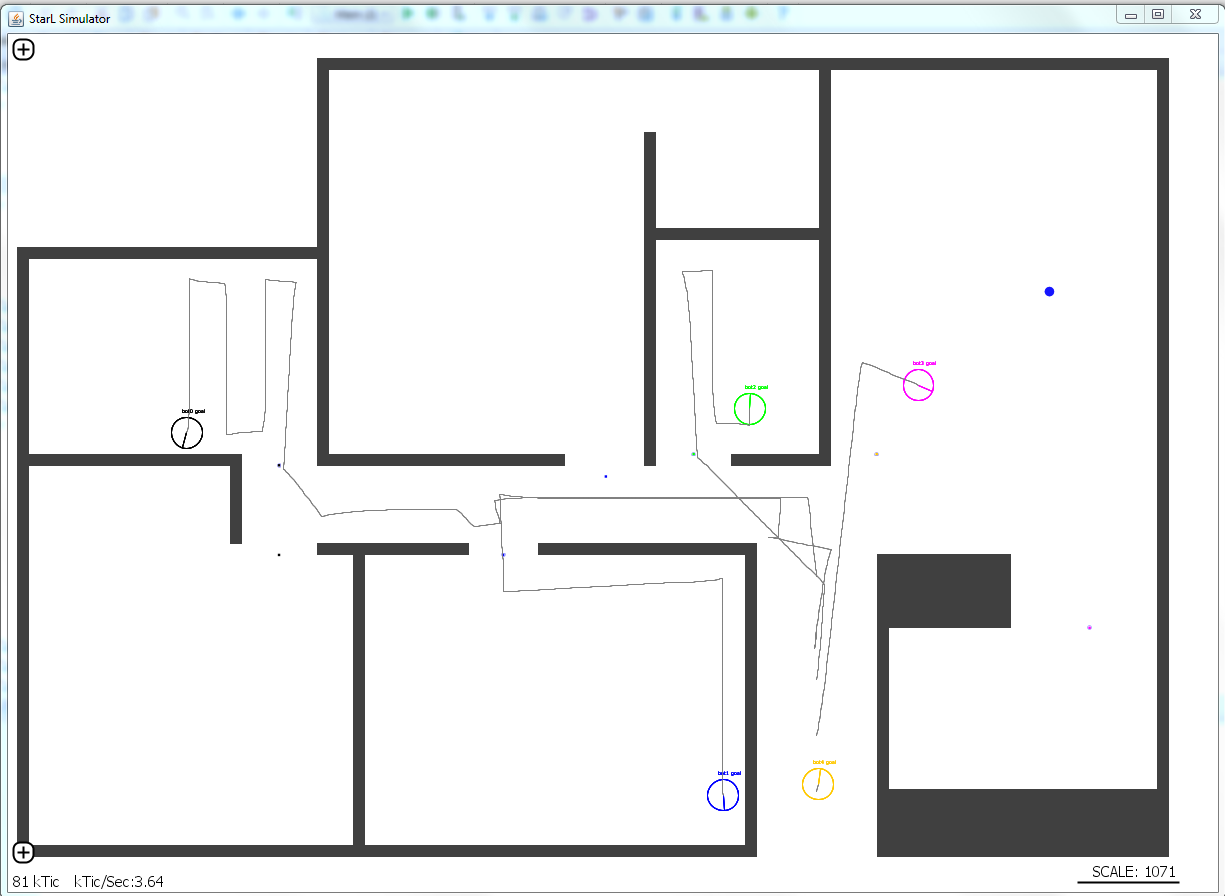}}
\caption{A simulation of the distributed search application.}
\label{fig:sim2}
\end{figure}

\begin{figure}
\centerline{\includegraphics[scale=0.35]{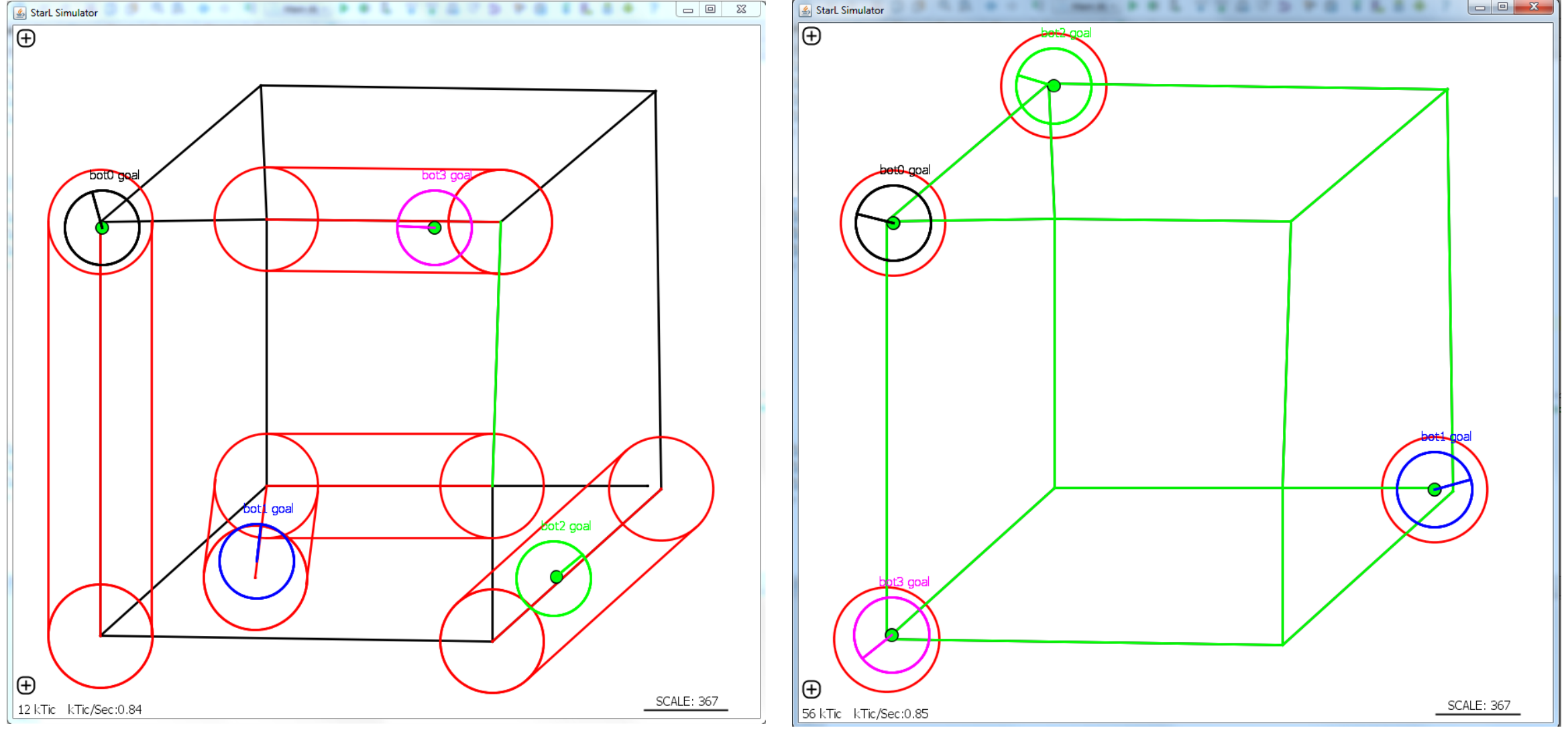}}
\caption{The light painting simulation.}
\label{fig:sim3}
\end{figure}

\subsection{Other Applications}
\label{sec:apps}

There are four other demo applications in StarL, including Race App, Maze App, Distributed Search App, Light Painting App. Each of them demonstrates some aspects of the StarL primitives.

In Race App, there is a sequence of destination points. Every robot picks the same destination point and tries to reach it before any other robot does. When robots collide with each other, they stop and turn until they are facing away from each other. When a robot reaches one destination point, it announces that through \starl{Broadcast} so that every robots starts to race to the next destination point. This application demonstrates how to make use of the motion and communication interface.

In Maze App, robots are put into a Maze like environment, which contains obstacles shown to robots and obstacles hidden to robots. The robot's goal is to navigate through the maze thus reaching the destination point. The robot uses the path planning primitive to find a possible path to the destination. When the robot's bump sensor detects an unseen obstacle, the robot updates its' obstacle map and recalculates path to the destination. This application demonstrates the path planning primitive as well as different built-in motion automation.

In Distributed Search App, a group of robots search a house to find an item. They first start the leader election primitive to elect a leader. Then the leader assigns rooms to each robot. Each robot goes to its assigned rooms and searches for the item. If the item is found, the robot announces it's finding to the group. A simulation screen shot is shown in figure\~ref{fig:sim2}. The item to be found is at the top right corner shown in a blue circle. Thin gray lines are robots' movement traces. The first three robots have entered their assigned room and started searching, the pink robot is moving towards it's assigned room at the bottom right corner. The yellow robot is still waiting for its assignment from the leader.

In Light Painting App, a simple diagram is given to a group of robots. The robots will try to plan their path to paint the lines in the diagram, with one or more colors. This application makes use of the distributed path planning primitive. A simulation for drawing a cube is shown in figure\~ref{fig:sim3}. The red tube is the distributed path planning reach tube for each robot. The painted lines are shown in green. On the left, the robots started to paint; on the right, the robots have finished the painting.

\section{Related Work}
\label{sec:related}
\paragraph{Robotic systems and theory}
There is a large body of theoretical work spanning control theory, computer science, and robotics that deals with development of distributed 
algorithms for flocking, coverage, and formation control for robotic swarms~\cite{DefagoK02,SY99:SIAM,prencipe01corda,MurrayOlfati,Magnusbook2010,coverage:IEEErobotics04,SchwagerMR06}. The safety and convergence properties of algorithms are typically analyzed by hand (as opposed to verified with a computer), under various simplifying assumptions. Control theorists and roboticists typically capture the details of the dynamics of the robots and abstract away the communication delays and issues arising from asynchrony, while the computer scientists make the complementary assumptions.

In the last ten years, these algorithms have been used to create spectacular robotic systems~\cite{kivaFORBES,michael2010grasp} and demonstrations~\cite{oung2011distributed,lindsey2012construction} 
for SLAM, flocking, collaborative search, and even construction.
In building these systems, each group uses its own specific, and often proprietary hardware and software architecture to implement the algorithms, with limited scope for reuse and no support for formal reasoning. In fact, currently there are no frameworks or tools supporting modular design, implementation, and formal verification of distributed robotic systems. 

\paragraph{Programming languages}
Currently robotic systems are programmed using standard programming languages like C, C++, and Python. It is also common to design low-level controllers using MATLAB/Simulink and then automatically generate C-code. 
The Robot Operating System (ROS)~\cite{ROSicra09} provides a popular set of libraries for building applications. It provides device drivers, message-passing and other low-level libraries for interfacing with sensors and actuators, and therefore, it could be used to build the lower layers in a StarL application.
Several synchronous programming languages like Lustre~\cite{Halbwachs91thesynchronous}, Esterel~\cite{Esterel94}, Signal~\cite{le1986signal}, and the Time-triggered framework~\cite{DBLP:conf/nfm/SteinerD11} have been developed over the past two decades. These languages are not only used in practice for signal processing, automotive, aerospace, and manufacturing applications, but they also provide strong formal semantics and support for verification. However, they all provide a deterministic programming abstraction and in one way or another we found them to be too restrictive for distributed robotic systems that work in highly dynamic environments.

\paragraph{Formal verification for distributed systems}
There is a large body of work on formal models for distributed systems or communicating state machines. A very general framework with limited verification support through PVS is the hybrid I/O automaton framework~\cite{TIOAmon,Mitra07PhD}. Differential dynamic logic~\cite{DBLP:conf/hybrid/Platzer07} with the related Keymera theorem prover is another well-developed framework. There are several less expressive models that have been developed for completely automatic verification under the umbrella of parameterized verification (see, for example,~\cite{abdulla1998verifying,DBLP:conf/cav/KrcalY06,JM:ICCPS:2012} and the references therein).

\section{Conclusions}
\label{sec:conc}
We presented what is to our knowledge a design of the first programming framework for distributed robotic systems that also supports simulations and rigorous verification. 
Since a robotic system is essentially an open system with many sources of nondeterminism, our primitives sacrifice determinism in the programming abstraction and instead provide a uniform way of interacting with physical environment, communication channels and other programs. 
The proposed StarL framework also provides theory libraries for verifying StarL applications in the PVS theorem prover, and 
two execution environments: one that is used to deploy the applications on smart phones that control robots, and 
the other for running discrete event simulations with many participating robots. 
The capabilities are illustrated with a StarL application for vehicle to vehicle coordination in a automatic intersection
that uses StarL primitives for point-to-point motion, mutual exclusion, and registration. 

The future directions of research include expansion of the StarL-PVS library further to include failure models and to support the verification of progress properties. 
Another direction is to develop a compiler for generating both the PVS theories (Figure~\ref{fig:pvs}) and the Java implementation (Figure~\ref{fig:java}) from the StarL programs (Figure~\ref{fig:gen_alg}). 

\section*{Acknowledgments}

We thank Adam Zimmerman for developing and documenting an earlier version of StarL development for his masters thesis research and Nitin Vaidya for several valuable discussions. This work is sponsored in part by the National Science Foundation.

\bibliographystyle{abbrv}
\bibliography{paper,sayan1}
\end{document}